\documentclass[aps,preprint,a4paper,showpacs,pre]{revtex4-1}
\usepackage{amsmath}
\usepackage{gensymb}
\usepackage{epsfig}
\usepackage{graphicx}
\usepackage{amsfonts, amssymb, upgreek}
\usepackage{textcomp}

\providecommand{\abs}[1]{\vert #1\vert}

\providecommand{\sigp}[1]{\hspace{.1ex}\sigma_{{#1}_+}}
\providecommand{\sigm}[1]{\hspace{.1ex}\sigma_{{#1}_-}}
\providecommand{\sigpm}[1]{\hspace{.1ex}\sigma_{{#1}_\pm}}

\newcommand{\Pst}{P\!_\mathrm{st}}
\newcommand{\sgn}{\mathop{}\!\mathrm{sgn}}
\newcommand{\supp}{\mathop{}\!\mathrm{supp}}

\newcommand{\upd}{\mathop{}\!\mathrm{d}}
\newcommand{\bsigma}{\boldsymbol{\sigma}}
\newcommand{\bvarrho}{\boldsymbol{\varrho}}
\newcommand{\balpha}{\boldsymbol{\alpha}}
\newcommand{\Or}{\mathrm{O}}

\begin{document}


\title{Coarse grained approach for universality classification of discrete models}
\author{R. C. Buceta}
\email{rbuceta@mdp.edu.ar}
\author{D. Hansmann}
\email{David.Hansmann@gmx.net}
\affiliation{Instituto de Investigaciones F\'{\i}sicas de Mar del Plata, Universidad Nacional de Mar del Plata and Consejo Nacional de Investigaciones Cient\'{\i}ficas y T\'ecnicas, Funes 3350, B7602AYL Mar del Plata, Argentina}

\begin{abstract}
Discrete and continuous models belonging to a universality class share the same linearities and (or) nonlinearities. 
In this work, we propose a new approach to calculate coarse grained coefficients of the continuous differential equation from discrete models. We apply small constant translations in a test space and show how to obtain these coefficients from the transformed average interface growth velocity. Using the examples of the ballistic deposition (BD) model and the restricted solid-on-solid (RSOS) model, both belonging to the Kardar-Parisi-Zhang (KPZ) universality class, we demonstrate how to apply our approach to calculate analytically the corresponding coefficients of the KPZ equation. Our analytical nonlinear coefficients are in agreement with numerical results obtained by Monte Carlo tilted simulations. In addition to the BD and the RSOS we study a competitive RSOS model that shows crossover between the KPZ and Edwards-Wilkinson universality classes.
\end{abstract}
\maketitle

\section{\label{sec:intro}Introduction}

Discrete and continuous models belonging to the same universality class are usually characterized by a unique set of power-law exponents related to each other by scaling relations \cite{Family1985, Family1986, Family1988, Barabasi1995}. Beyond that, discrete and continuous models within the same class share the same linearities and (or) nonlinearities \cite{Lai1991}. All this properties are result of the evolution rules determined by the transition rates between surface configurations. Usually these rules are based on nested \texttt{IF-THEN-ELSE} structures, which alternatively can be expressed by means of Heaviside unit-step functions $\Theta$. Starting from the Master equation of the configuration probability, it is possible to derive an associated discrete Langevin equation performing the Kramers-Moyal expansion \cite{Kramers-40, Moyal-49,Fox-78,VanKampen-81}.
Note that the \mbox{$\Theta\,$-function} is a distribution or generalized function and therefore the drift function of the discrete Langevin equation is a distribution, too. The continuous Langevin equation can be achieved from its discrete counterpart following the approach introduced by Vvedensky {\sl et al.} \cite{Vvedensky-93}. 
This approach is based on regularization techniques and on the coarse grain approximation, consistent with the limit of lattice constant tending to zero \cite{Park1995, Predota1996, Oliveira2006, Costanza1997, Costanza2009, Buceta2005, Muraca2004}. The regularization method involves the replacement of the Heavisides $\Theta$ by smooth functions $\theta_\varepsilon$. Here the regularizing parameter $\varepsilon$ has to be chosen in a way that $\theta_\varepsilon\to\Theta$ when $\varepsilon\to 0^+$.  The coarse grain approximation is carried out by a Taylor expansion around the origin of either the entire regularized drift function or each single regularized function $\theta_\varepsilon$ and its arguments ({\sl e.g.} Laplacian or gradient). It can be observed, that the results of these two expansion methods are ambiguous, since they strongly depend on the chosen regularization \cite{Predota1996}.\newline
\indent Within the group of discrete models, which describe deposition or evaporation processes, the most studied ones belong to the Edwards-Wilkinson (EW) or the Kardar-Parisi-Zhang (KPZ) universality class. Models of the EW class treat slow deposition processes with relaxation under gravity \cite{Edwards1982}. Models of the KPZ class involve deposition (or evaporation) processes that permit lateral growth (or decrease) \cite{Kardar1986}. Usually studies of these models were focused on the characterization of statistical surface growing properties, {\sl e.g.} roughness and growth exponents of the surface width and scaling function in the steady state. The KPZ universality class includes several discrete models, {\sl e.g.} the restricted solid-on-solid (RSOS) model \cite{Kim-89}, the Eden model \cite{Eden-61, Plischke1984}, ballistic deposition (BD) model \cite{Family1985, Vold1963, Sutherland1966, Meakin1986}, and the deposition-evaporation model \cite{Plischke1987}. Among these models, many authors chose to study the RSOS model, as it shows a very good scaling behaviour even with small systems and in addition reaches rapidly the steady state. First, Park and Kahng \cite{Park1995} have derived the KPZ equation \cite{Kardar1986} from the RSOS model. More recently using the same method, Oliveira {\sl et al.} \cite{Oliveira2006} has derived the KPZ equation from a competitive restricted solid-on-solid (CRSOS) model involving deposition and evaporation, with probabilities $p$ and $1-p$, respectively. Other authors have derived the KPZ equation starting from the BD model following different schemes of regularizations \cite{Costanza1997, Costanza2009, Buceta2005}.\newline
\indent In this work we propose a new approach to derive the linear and (or) nonlinear coefficients of a corresponding continuous equation starting from a discrete model, and thus to classify discrete models. 
We introduce the concept of interface configuration space (ICS) in which discrete processes are defined. The components of a point in the ICS are the height differences between neighbour columns and the site column evolved. Within this configuration space the drift function of discrete Langevin equation and the stationary probability density function (SPDF) are defined. Additionally, we assume that the SPDF for a restricted (or unrestricted) discrete process is embedded in a space of test functions with compact support (or rapidly decreasing behaviour). We show that proper constant translations applied to the test function transform the average velocity of the interface growth. The coefficients that occur under the proposed transformations correspond to the coarse grained coefficients of continuous differential equation which defines the universality class. In the framework we focus on the RSOS and the BD model of the KPZ class, since they can be taken as a reference of a restricted and unrestricted processes respectively where our theoretical approach can be applied.   

\subsection{From Master equation to discrete Langevin equation} 
Let us consider a surface configuration $\mathbf{H}$, which is determined by a set of heights $\{h_j\}$ corresponding to the columns $j$, with $h_j\in\mathbb{Z}$. The transition rate $W(\mathbf{H},\mathbf{H}')$ between two surface configurations $\mathbf{H}$ and $\mathbf{H}'$, for a process that evolves (increasing or decreasing $r_k$-units) at the selected column $k$ is
\begin{equation}
W(\mathbf{H},\mathbf{H}')\!=\!\frac{1}{\tau}\sum_{k=1}^M\omega_k(\mathbf{H},\mathbf{H}')\Delta(h_k',h_k+r_k)\prod_{j\neq k}\Delta(h_j',h_j)\,,\nonumber
\end{equation}
where $M$ is the system size and $\Delta(x,y)$ is equal to $1$ if $x=y$ and equal to $0$ otherwise. Here the hopping rules $\omega_k(\mathbf{H},\mathbf{H}')$ take the restraints process into account. For restricted models, {\sl e.g.} the RSOS $\omega_k$ is function of the height difference between the selected column $k$ and its next neighbours. For unrestricted models, {\sl e.g.} the BD $\,\omega_k= 1$. The growth $r_k$ of the column $k$ depend on the underlying discrete growth model, e.g. $r_k=1$ for the RSOS model and $r_k=\max(\sigp{k},\sigm{k},1)$, with \mbox{$\sigpm{k}=h_{k\pm 1}-h_k\,$} for the BD model. The first and second transition moments are
\begin{eqnarray*}
&&\hspace{-7ex} K_j^{(1)}\bigl(\bsigma_j\bigr)\!=a\,\sum_{\mathbf{H}'}\;(h_j'-h_j)\,W(\mathbf{H},\mathbf{H}')\,,\\
&&\hspace{-7ex} K_{ij}^{(2)}\bigl(\bsigma_j\bigr)\!=a^2\,\sum_{\mathbf{H}'}\;(h_i'-h_i)(h_j'-h_j)\,W(\mathbf{H},\mathbf{H}')\;.
\end{eqnarray*}
where $\bsigma_j=(\sigma_{j1},\dots,\sigma_{jN})\in\mathbb{Z}^N$ and $\sigma_{jk}=h_k-h_j$ is the height difference between the neighbour column $k$ and the selected column $j$. Here $N$ is the number of nearest neighbours.
Using a Kramers-Moyal expansion from the Master equation is derived the following Fokker-Planck equation  [$P=P\bigl(\{\bsigma_j\},t\bigr)$]
\begin{equation}
\frac{\partial P}{\partial t}=-\frac{\partial\quad}{\partial h_j}(K_j^{(1)} P)+\frac{1}{2}\;\frac{\partial^2\qquad}{\partial h_i \partial h_j}(K_{ij}^{(2)} P)\;.\label{FP}
\end{equation}
The corresponding Langevin equation is ($\partial_t\doteq\partial/\partial t$)
\begin{equation}
\partial_t\,h_j=K_j\bigl(\bsigma_j)+\eta_j(t)\;,\label{Langevin}
\end{equation}
where $K_j\,\dot=\,K^{(1)}_j$ is the drift term and $\eta_j$ is Gaussian white noise with $\langle\eta_j\rangle=0$ 
and $\langle\eta_i(t)\eta_j(t')\rangle=Q_{ij}\,\delta(t-t')\;$.
Here $Q_{ij}\dot=K^{(2)}_{ij}$ is the noise intensity. Introducing the probability current ($\partial_j\doteq\partial/\partial h_j$)
\begin{equation}
J_i=K_i P-\tfrac{1}{2}\;\partial_j(Q_{ij} P)\;,\nonumber
\end{equation}
the Fokker-Planck equation~(\ref{FP}) is
\begin{equation}
\partial_t P=-\partial_i J_i\;.\nonumber
\end{equation}
In the steady state $\partial_t\Pst=0$, where $\Pst=\Pst(\bsigma_j)$ is the SPDF with the property 
$\sum_{\bsigma\in\mathcal{Z}^N}\Pst(\bsigma)=1\;$.

\subsection{\label{subsec:transf}Transformations on interface configuration}
The first approach that allows to characterize nonlinearities of discrete models was the surface tilt transformation \cite{Krug1989, Huse-90}.
It is based on a global transformation $\nabla h\to\nabla h+\mathbf{s}$ which is applied to the continuous KPZ equation [$h=h(\mathbf{x},t)$ with $\mathbf{x}\in\mathbb{R}^n$]
\begin{equation}
\frac{\partial h}{\partial t}=\nu\,\nabla^2 h +\frac{\lambda}{2}\,\abs{\nabla h}^2+\eta(\mathbf{x},t)\;,
\end{equation}
and transforms the average velocity of interface growth as\cite{Barabasi1995}
\begin{equation}
v(0)\to v(s)=v(0)+\frac{\lambda}{2}\,s^2\;,
\end{equation}
with $s=\abs{\mathbf{s}}$. Employing the tilt transformation and Monte Carlo techniques, several authors characterized nonlinearities of discrete models belonging to the KPZ universality class \cite{Huse-90, Krug-90, Amar-92}. In this work we propose direct transformations of the discrete Langevin equation (\ref{Langevin}). For simplicity we consider only two next neighbours $\bsigma_j\!=\!(\sigp{j},\sigm{j})$ and use the tilt transformation $h_j\to h_j+s j\;$ with $s\in\mathbb R$. In this case the height differences between the nearest-neighbours are transformed as $\sigpm{j}\to\sigpm{j}\pm s$. We define the two auxiliary variables 
\begin{eqnarray}
&&\eta_j\doteq\tfrac{1}{2}(\sigp{j}-\sigm{j})=\tfrac{1}{2}(h_{j+1}-h_{j-1})\;,\label{eta}\nonumber\\
&&\zeta_j\doteq\tfrac{1}{2}(\sigp{j}+\sigm{j})=\tfrac{1}{2}(h_{j+1}-2 h_j+h_{j-1})\nonumber\;.
\end{eqnarray}
These new variables are the standard (or post-point) discretization of the gradient \cite{Buceta2005} and the discretization of the Laplacian.
Using the tilt transformation, these variables change as $\eta_j\to\eta_j+s\;$ and $\;\zeta_j\to\zeta_j\;$ i.e. the transformation shifts the ``gradients'' and leaves the ``Laplacian'' invariant, respectively.
In contrast, the transformation $\sigpm{j}\to\sigpm{j}+ r$ leaves ``gradient'' invariant and changes the ``Laplacian'' to $\;\zeta_j\to\zeta_j+r$. 
These transformations may be interpreted as translations along the ``gradient'' and ``Laplacian'' directions of the surface configuration space. 

\section{\label{sec:distributions}Discrete processes revisited from the theory of distributions}
Let $\mathcal{Z}^N\!=\bigl\{\bsigma_j\in\mathbb{Z}^N/\Pst(\bsigma_j)\neq 0\,,\,\forall j\}\,$ be the set of surface configurations in the steady state. Here $\mathcal{Z}^N$, called interface configuration space (ICS), is the $N$-dimensional lattice in the compact set \mbox{$\mathcal{R}^N\subset\bigl(\mathbb{R}^N\setminus\mathbb{Z}^N\bigr)\cup\mathcal{Z}^N$}, consisting of vectors $\bsigma_j$ with integer coordinates \mbox{$\sigma_{jk}=h_k-h_j$} ($k=1,\dots ,N$). The drift term of the discrete Langevin eq.~(\ref{Langevin}) can be assumed as a generalized function $K(\bvarrho)$ with $\bvarrho\in\mathbb{R}^N$ evaluated at \mbox{$\bvarrho\!=\!\bsigma_j\in\mathbb{Z}^N$}. 
In order to calculate statistical observables of restricted and not restricted discrete processes, it is needed to define test functions $\varphi$ on which the generalized functions of $\bvarrho$ are applied. The restricted processes require that \mbox{$\varphi\in\mathcal{D}(\mathcal{R}^N_\varepsilon)$}. 
Here $\mathcal{D}$ is the test space of $\mathcal{C}^\infty$-functions with compact support\footnote{The support of the function \mbox{$\varphi:{\mathcal{R}}^N_\varepsilon\to\mathbb{C}$}, denoted $\supp(\varphi)$, is the closure of \mbox{$\{\bvarrho\in {\mathcal{R}}^N_\varepsilon/\varphi(\bvarrho)\neq 0\}$}.} in  \mbox{$\mathcal{R}^N_\varepsilon= \varepsilon-\rm{neighbourhood}(\mathcal{R}^N)\,=\{\bvarrho\in\mathbb{R}^N/d(\bvarrho,\mathcal{R}^N)<\varepsilon\}\,$}.  The compact support excludes interface configurations in $\mathbb{Z}^N\setminus \mathcal{Z}^N$ since $\supp(\varphi)\subset\mathcal{R}^N_\varepsilon$. In contrast, unrestricted processes require that $\varphi\in\mathcal{S}(\mathbb R^N)$. Here $\mathcal{S}$ is the test space of $\mathcal{C}^\infty$-functions that decay and have derivatives of all orders that vanish faster than any power of $\varrho_\alpha^{-1}$ ($\alpha=1,\dots,N$). Our notation about distributions is the standard, first established by Schwartz \cite{Schwartz-66}. 
The application of a distribution $f\in\mathcal{D}'$ (dual space of $\mathcal{D}$) to test function $\varphi\in \mathcal{D}$ is defined by 
\begin{equation}
\langle f\,,\varphi\rangle=\int_{\mathbb{R}^N}f(\bvarrho)\,\varphi(\bvarrho)\;\upd^N\!\varrho\;.\label{f}
\end{equation}
Please note that $\langle f,\varphi\rangle$ is the expectation value of $f$ using the test function $\varphi$ as real analytical ``representation'' of the SPDF $\Pst$. Thus, the test function is normed, {\sl i.e.} $\langle 1,\varphi\rangle=1\,$.
The translation $T_{\balpha}$ of a distribution $f$, denoted $T_{\balpha} f$ or simply $f_{\balpha}$,
extends the definition given by eq.(\ref{f}) to
\begin{equation}
\langle T_{\balpha} f\,,\varphi\rangle=\int_{\mathbb{R}^N}f(\bvarrho-\balpha)\,\varphi(\bvarrho)\;\upd^N\!\varrho\;,\label{Tx}
\end{equation}
where the translation operator is defined by \mbox{$T_{\mathbf{x}}:\mathbf{y}\mapsto\mathbf{y}-\mathbf{x}$} if $\mathbf{y}\,,\mathbf{x}\in\mathbb{R}^N$ \cite{Colombeau1984}. Notice that eq.~(\ref{Tx}) satisfy
\begin{equation}
\langle T_{\balpha} f\,,\varphi\rangle=\langle f\,, T_{-\balpha}\,\varphi\rangle\,.\label{T}
\end{equation}
We assume that the test function $\varphi$ takes ​​fixed values ​​in the configuration space $\mathcal{Z}^N$ given by the SPDF $\Pst$, {\sl i.e.} 
\begin{equation*}
\langle \delta_{\bsigma}\,,\varphi\rangle=\varphi(\bsigma)=\Pst(\bsigma)\,,\hspace{2cm}\forall\,\bsigma\in\mathcal{Z}^N\,,
\end{equation*}
where $\delta_{\bsigma}=T_{\bsigma}\delta$ and $\delta$ is the Dirac distribution. 

\subsection{Transformations on the average velocity}

Applying a translation $T_{\mathbf{u}}$ to a point $\bvarrho\in\supp(\varphi)$, the test function transforms as $\varphi\rightarrow T_{\mathbf{u}}\varphi$\, if\,  $(\bvarrho-\mathbf{u})\in\supp(\varphi)$ and the average velocity of the growth interface change as $v(0)\rightarrow v(\mathbf{u})$ with
\begin{equation}
v(\mathbf{u})=\langle K\,,T_{\mathbf{u}}\varphi\rangle=\int_{\mathbb{R}^N}K(\bvarrho)\;\varphi(\bvarrho-\mathbf{u})\;\upd^N\!\varrho\;.\label{drift}
\end{equation}
For small translations the Taylor expansion of $\varphi(\bvarrho-\mathbf{u})$ around $\mathbf{u}=\mathbf 0\,$ is
\begin{equation}
\varphi(\bvarrho-\mathbf{u})=\varphi(\bvarrho)-\partial_\alpha\varphi\bigr\rfloor_{\mathbf{u}=0}\;u_\alpha+\tfrac{1}{2}\,\partial^2_{\alpha\beta}\varphi\bigr\rfloor_{\mathbf{u}=0}\;u_\alpha u_\beta+\Or(3)\;.\label{Tvarphi}
\end{equation}
Here repeated subscripts imply sums, $u_\alpha$ is the $\alpha$-th component of $\mathbf{u}$, $\partial_\alpha\,\dot=\,\partial/\partial\varrho_\alpha$, and $\partial^2_{\alpha\beta}\,\dot=\,\partial^2/(\partial\varrho_\alpha\partial\varrho_\beta)$, with $\alpha=\beta=1,\dots,N$. 
Since the test function $\varphi$ is known only at points of the lattice $\mathcal{Z}^N$, its derivatives can not be calculated explicitly. In contrast, the drift distribution $K$ is derivable in all points. Since the test function either has compact support or decreases rapidly one can take advantage of the following identity 
\begin{equation}
\bigl\langle K\,,\partial^{\rm{n}}_{\alpha\beta\cdots\omega}\varphi\bigr\rangle=(-1)^{\rm{n}}\bigl\langle \partial^{\rm{n}}_{\alpha\beta\cdots\omega} K\,,\varphi\bigr\rangle\;.\label{repl}
\end{equation}
Using eqs.~(\ref{drift}--\ref{repl}), the average velocity is transformed according to
\begin{equation}
v(\mathbf{u})=\Bigl\langle K,\varphi\Bigr\rangle +\Bigl\langle\partial_\alpha K,\varphi\Bigr\rangle\,u_\alpha
+\tfrac{1}{2}\Bigl\langle\partial^2_{\alpha\beta} K,\varphi\Bigr\rangle\,u_\alpha u_\beta+\Or(3)\,.\label{distr-exp0}
\end{equation}
The physical meaning of the translations in test space can be made clear considering the case $N\!=\!2$. It corresponds to a discrete process that describes the evolution of an interface in (1+1)--dimensions. $K(\varrho_1,\varrho_2)$ is symmetric under the exchange $\varrho_1\leftrightarrow\varrho_2$, for an isotropic process. Consequently $\varphi(\varrho_1,\varrho_2)$ is symmetric under this exchange, too. Considering a translation $\mathbf{u}$ in the $\supp(\varphi)$ with components $u_1=r+s$ and $u_2=r-s$, the non-zero lower orders in the Taylor series expansion are
\begin{equation}
v(r,s)\simeq \Bigl\langle K\,,\varphi\Bigr\rangle+\Bigl\langle(\partial_1+\partial_2) K\,,\varphi\Bigr\rangle\;r+\;\tfrac{1}{2}\,\Bigl\langle(\partial^2_{11}-2\,\partial^2_{12}+\partial^2_{22}) K\,,\varphi\Bigr\rangle\;s^2\;.\label{distr-exp}
\end{equation}
Here the linear term in $s$ vanishes because $(\partial_1-\partial_2)K$ is antisymmetric. Taking into account that   $(\partial_{1}+\partial_{2})K$ and $(\partial^2_{11}-2\,\partial^2_{12}+\partial^2_{22})K$ are symmetric eq.~(\ref{distr-exp}) is
\begin{equation}
v(r,s)\simeq v_0+\nu\;r+\tfrac{1}{2}\,\lambda\,s^2\;,\label{speed-rs}
\end{equation} 
with the coefficients ($\alpha,\beta=1,2$ with $\alpha\neq\beta$)
\begin{eqnarray}
\nu&=&2\,\Bigl\langle\partial_\alpha K\,,\varphi\Bigr\rangle\;,\label{nu}\\
\lambda&=&2\,\Bigl\langle\bigl(\partial^2_{\alpha\alpha}-\partial^2_{\alpha\beta}\bigr) K\,,\varphi\Bigr\rangle\;,\label{lambda}\\
v_0&=&\Bigl\langle K\,,\varphi\Bigr\rangle=v(0,0)\;,\label{v0}
\end{eqnarray}
where repeated subscripts do not imply sums. Assuming that higher-order nonlinearities can be neglected,
discrete models belong to KPZ universality class ($\lambda\neq 0$) if $\bigl(\partial^2_{\alpha\alpha}-\partial^2_{\alpha\beta}\bigr) K$ is not zero and symmetric under the exchange $\varrho_1\leftrightarrow\varrho_2\,$. Otherwise, the discrete models belong to Edwards-Wilkinson (EW) universality class ($\lambda=0$).
Equation~(\ref{speed-rs}) can alternatively be obtained applying translations to the gradient and the Laplacian term of the KPZ equation, {\sl i.e.} $\nabla^2 h\to\nabla^2 h+r$ and $\nabla h\to\nabla h+s$, and then taking averages over the interface. The set of parameters $\{\nu,\lambda\}$ defined by eqs.~(\ref{nu}) and (\ref{lambda}) is unique for each isotropic discrete model. 

\section{\label{sec:RSOS}RSOS Model}
\begin{figure}[t!]
\centerline{\includegraphics[width=0.5\linewidth]{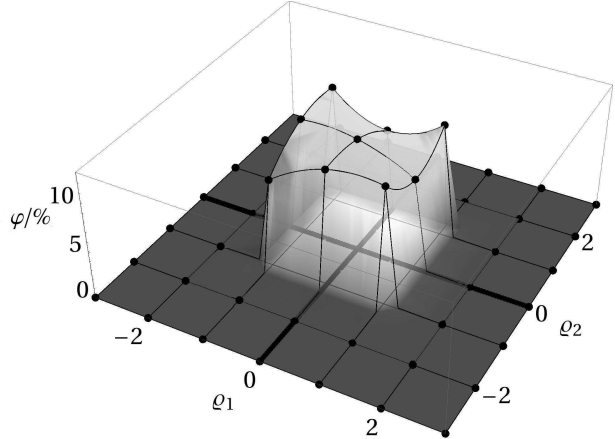}}
\caption{\footnotesize For the RSOS model we plot a smooth test function $\varphi$ as function of $(\varrho_1,\varrho_2)$. The non zero values of the SPDF (bullets) obtained by Monte Carlo simulation are: $\varphi(0,0)=\varphi(1,1)=0.1158,\;\varphi(0,1)=\varphi(1,0)=0.0944,\;\varphi(-1,1)=\varphi(1,-1)=0.1249,\;\varphi(-1,0)=\varphi(0,-1)=0.1198,\;\rm{and}\;\varphi(-1,-1)=0.0903$. See that $\supp(\varphi)$ is the projection of $\varphi\neq 0$ on the plane (light gray).}\label{fig1}
\end{figure}

For a single-step aggregation process in one dimensional substratum, the RSOS condition is $\abs{\varrho_k}\le 1$ with $k=1,2\,$. Evolving from a flat interface we observe 9 interface configurations defined by $\mathcal{Z}^2=\{(i,j)\in\mathbb{Z}^2/\,i,j=-1,0,1\}$. The test function $\varphi(\varrho_1,\varrho_2)$ is taken as a $\mathcal{C}^\infty$-function with compact support, defined in $\mathcal{R}^2_\varepsilon=\varepsilon-\rm{neighbourhood}(\mathcal{R}^2)$ an open subset of $\mathbb{R}^2$ that excludes all points of $\mathbb{Z}^2\setminus\mathcal{Z}^2$\,. For simplicity we take $\mathcal{R}^2=\{\bvarrho\in\mathbb{R}^2/\abs{\varrho_k}\le 1,\; k=1,2\}$ (see fig.~\ref{fig1}). The drift of Langevin eq.~(\ref{Langevin}) is the generalized function
\begin{equation}
K(\varrho_1,\varrho_2)=\Theta(\varrho_1)\,\Theta(\varrho_2)\;,
\end{equation}
where we take $a/\tau=1$ without loss of generality and the Heaviside function $\Theta(z)$ is equal to $1$ if $z\ge 0$ and equal to $0$ otherwise. The derivatives of the $K$ up to 2-nd order are ($\alpha,\beta=1,2$ and $\alpha\neq\beta$)
\begin{eqnarray}
\partial_\alpha K &=& \delta(\varrho_\alpha)\,\Theta(\varrho_\beta)\,,\nonumber\\
\partial^2_{\alpha\beta} K&=&\delta(\varrho_\alpha)\,\delta(\varrho_\beta)\,,\\\
\partial^2_{\alpha\alpha} K&=&\delta'(\varrho_\alpha)\,\Theta(\varrho_\beta)\,.\nonumber
\end{eqnarray}
Applying these distributions to $\varphi$ we obtain the coefficients $\nu$ and $\lambda$ defined by eqs.~(\ref{nu}) and (\ref{lambda}), respectively. Explicitly
\begin{eqnarray}
\hspace{-8ex}&&\nu=2\,\int_0^{1+\epsilon}\varphi(0,y)\,\upd y\,,\label{nu-rsos}\\
\hspace{-8ex}&&\lambda=-2\,\left[\varphi(0,0)+\int_0^{1+\epsilon}\partial_x\varphi\Bigr\rfloor_{(0,y)}\,\upd y\right]\,.\label{lambda-rsos}
\end{eqnarray}
\paragraph*{Numerical results:} In this work, the SPDF were calculated {\sl via} Monte Carlo simulations using system sizes $M=1024$, with periodic boundary conditions, at $t>10^3$ monolayers, taking the average of $10^3$ samples. In order to evaluate eqs.~(\ref{nu-rsos}) and (\ref{lambda-rsos}) we take $\epsilon\to 0^+$ and assume that $\varphi$ and $\partial_x\varphi$ at $x=0$ are very smooth functions of $y$ in $(0,1)$. Straightforward we obtain $\nu\simeq\varphi(0,0)+\varphi(0,1)$.
We obtain $\lambda\simeq -2\,(\varphi+\partial_x\varphi+\tfrac{1}{2}\,\partial^2_{xy}\varphi)\rfloor_{(0,0)}\,$
replacing $\partial_x\varphi\rfloor_{(0,y)}\simeq\partial_x\varphi\rfloor_{(0,0)}+y\,\partial^2_{xy}\varphi\rfloor_{(0,0)}$ in eq.~(\ref{lambda-rsos}).
Please note that $\partial_x\varphi\rfloor_{(0,0)}\!\simeq\![\varphi(1,0)\!-\!\varphi(-1,0)]/2$, and $\partial_{xy}\varphi\rfloor_{(0,0)}\simeq-[\varphi(1,1)-2\,\varphi(-1,1)+\varphi(-1,-1)]/4$. With the values given in the caption of fig.~\ref{fig1} we obtain $\nu\simeq0.2102\,$ $\lambda\simeq -0.2171$.

\section{\label{sec:CRSOS}CRSOS Model}
The simplest competitive model that shows a crossover between EW and KPZ universality classes is the deposition\--evaporation model with RSOS condition. The deposition (or evaporation) happens with probability $p$ (or $1-p$), in a one dimensional substratum. This model was first numerically studied by Amar and Family far from crossover \cite{Amar-92}. Later Olivera {\sl et al.} showed analytically the $\lambda$-$p$ relation at the associated KPZ equation \cite{Oliveira2006}. Both,
the interface configuration space $\mathcal{Z}^2$ and test space $\mathcal{R}^2$ are identically to the ones of the RSOS model. The drift term of the CRSOS model is
\begin{equation}
K(\varrho_1,\varrho_2)=p\,\Theta(\varrho_1)\,\Theta(\varrho_2)-(1-p)\,\Theta(-\varrho_1)\,\Theta(-\varrho_2)\;,\label{K-crsos}
\end{equation} 
and its derivatives up to 2-nd order are
\begin{eqnarray}
\partial_\alpha K&=&\delta(\varrho_\alpha)\bigl[p\,\Theta(\varrho_\beta)+(1-p)\,\Theta(-\varrho_\beta)\bigl]\,,\nonumber\\
\partial_{\alpha\beta} K&=&(2\,p-1)\,\delta(\varrho_\alpha)\,\delta(\varrho_\beta)\,,\\\
\partial_{\alpha\alpha} K&=&\delta'(\varrho_\alpha)\bigl[p\,\Theta(\varrho_\beta)+(1-p)\,\Theta(-\varrho_\beta)\bigl]\,.\nonumber
\end{eqnarray}
Applying these distributions to $\varphi$ we obtain the KPZ coefficients defined by eqs.~(\ref{nu}) and (\ref{lambda}). 
The nonlinear coefficient is \footnote{Use $p\,\Theta(y)+(1-p)\Theta(-y)=\frac{1}{2}(2p-1)\sgn(y)+\frac{1}{2}[1+\Delta(y,0)]$.} 
\begin{equation}
\lambda=-(2 p-1)\left[2\,\varphi(0,0)+\int_{-1-\epsilon}^{1+\epsilon}\sgn(y)\,\partial_x\varphi\Bigr\rfloor_{(0,y)}\upd y\right]-\int_{-1-\epsilon}^{1+\epsilon}\partial_x\varphi\Bigr\rfloor_{(0,y)}\upd y\;.\label{lambda-crsos}
\end{equation}
Thus, when $p=1/2$ the process belongs to the EW universality class coinciding with the results obtained by other approaches \cite{Oliveira2006,Lazarides2006}. Please note that if $p=1/2$ then $\partial_x\varphi\rfloor_{(0,y)}$ is odd function of $y$, and consequently $\lambda=0$. To understand the latter take into account that the test function satisfies $\varphi(x,y)=\varphi(-x,-y)$, because the competitive processes is equiprobable. When this symmetry breaks the KPZ universality class appears.

\section{\label{sec:BDM}Ballistic Deposition Model}

\begin{figure}[t]
\centerline{\includegraphics[width=0.5\linewidth]{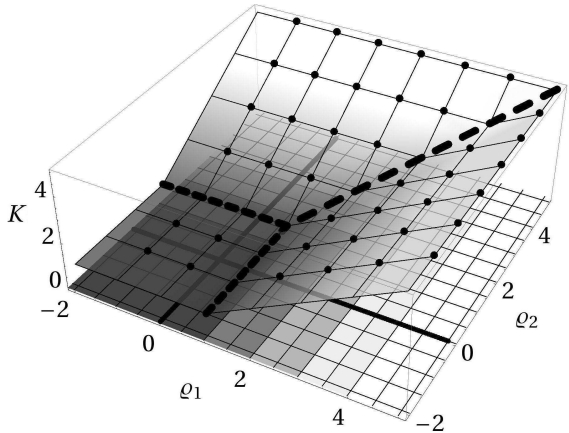}}
\caption{\footnotesize For the BD model we plot the drift distribution $K$ as function of $(\varrho_1,\varrho_2)$, given by eq.~(\ref{K-BD}). The bullets show the discrete drift $K(i,j)$ where $(i,j)\in\mathbb{Z}^2$. Dashed line shows where the distribution is not smooth.}
\label{fig2}
\end{figure}
The evolution of the ballistic deposition (BD) model in (1+1)-dimensions is defined for all $(\varrho_1,\varrho_2)\in\mathbb{R}^2$ by the distribution
\begin{eqnarray}
K(\varrho_1,\varrho_2)=\max(\varrho_1,\varrho_2,1)\,,\label{K-BD}
\end{eqnarray} 
where we define $a/\tau=1$ without loss of generality. Alternatively the BD evolution rules can be written as
\begin{equation}
K=g\,\Theta(g-1)+[1-\Theta(g-1)]\;,
\end{equation}
with
\begin{equation}
g=\max(\varrho_1,\varrho_2)=\zeta+\abs{\eta}\,.
\end{equation}
Here the auxiliary variables $\zeta$ and $\eta$ are again defined as
$\zeta=\frac{1}{2}(\varrho_1+\varrho_2)$ and $\eta=\frac{1}{2}(\varrho_1-\varrho_2)$. The 2-nd order Taylor expansion of $K(\zeta+r,\eta+s)$ around $r=0$ and $s=0$ is
\begin{equation}
K(\zeta+r,\eta+s) \simeq\,K(\zeta,\eta)+(r\partial_\zeta+s\partial_\eta) K+\;\tfrac{1}{2}\,(r^2\partial^2_{\zeta\zeta}-2\,r s\, \partial^2_{\zeta\eta}+s^2\partial^2_{\eta\eta}) K\;.\label{K-rs}
\end{equation}
Taking into account that $\partial_\eta g=\abs{\eta}'=\sgn(\eta)=\sgn(\varrho_1-\varrho_2)$ \cite{Kanwal-83} and $\partial^2_{\eta\eta}g=2\,\delta(\eta)=4\,\delta(\varrho_1-\varrho_2)$, where $\sgn(\eta)=\Theta(\eta)-\Theta(-\eta)$ is the sign function, one achieves
\begin{eqnarray}
\partial_\zeta K&=&(g-1)\,\delta(g-1)+\Theta(g-1)\,,\nonumber\\
\partial_\eta K &=& \Bigl[(g-1)\,\delta(g-1)+\Theta(g-1)\Bigr]\,\sgn(\eta)\,,\nonumber\\
\partial^2_{\zeta\zeta} K &=& \partial_\zeta\Bigl[(g-1)\,\delta(g-1)\Bigr] + \delta(g-1)\,,\nonumber\\
\partial^2_{\zeta\eta} K &=& \partial_\eta\Bigl[(g-1)\,\delta(g-1)\Bigr] + \delta(g-1)\,\sgn(\eta)\,,\label{d-ez}\\
\partial^2_{\eta\eta} K &=& 2\,\Bigl[(g-1)\,\delta(g-1)+\Theta(g-1)\Bigr]\,\delta(\eta)\nonumber\\
&&+\,\partial_\eta\Bigl[(g-1)\,\delta(g-1)\Bigr]\,\sgn(\eta)+\,\delta(g-1)\,[\sgn(\eta)]^2\,.\nonumber
\end{eqnarray}
Applying the distribution [eqs.~(\ref{K-rs})] to the test function $\varphi$ one obtains eq.~(\ref{speed-rs}) with
\begin{eqnarray}
\nu \!&=&\!\Bigl\langle\partial_\zeta K\,,\varphi\Bigr\rangle = \Bigl\langle\Theta(g-1)\,,\varphi\Bigr\rangle\,,\nonumber\\
\lambda\!&=&\! \Bigl\langle\partial^2_{\eta\eta} K\,,\varphi\Bigr\rangle = 4\,\Bigl\langle\Theta(g-1)\,\delta(\varrho_1-\varrho_2)\,,\varphi\Bigr\rangle\nonumber\\&&\hspace{14ex}+\Bigl\langle\delta(g-1)\,[1-\bigtriangleup(\varrho_1,\varrho_2)]\,,\varphi\Bigr\rangle\,,\nonumber\\
v_0 \!&=&\! \Bigl\langle K\,,\varphi\Bigr\rangle = 1-\nu+\Bigl\langle g\,\Theta(g-1)\,,\varphi\Bigr\rangle\,.\label{bm-rs}
\end{eqnarray}
Here we state only non zero contributions and $\bigtriangleup(\varrho_1,\varrho_2)=1-[\sgn(\varrho_1-\varrho_2)]^2$. Note that the application of the distribution $(g-1)\,\delta(g-1)$ (or its derivatives) on a test function is always zero. Also, the antisymmetric sign distribution applied to a symmetric test function is zero. The term of 2-nd order in $r$ has non-zero coefficient $\bigl\langle\delta(g-1)\,,\varphi\bigr\rangle$, which is negligible compared to $\nu$.
\begin{figure}[t!]
\centerline{\includegraphics[width=0.5\linewidth]{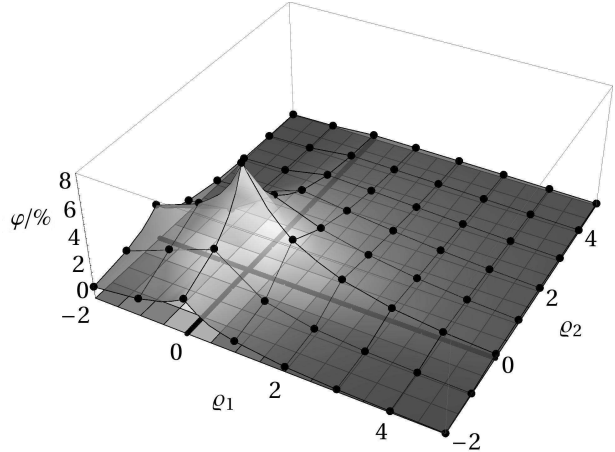}}
\caption{\footnotesize For the BD model we plot a decreasing test function $\varphi$ as function of $(\varrho_1,\varrho_2)$,  which take fixed values $\varphi(i,j)$ obtained from the Monte Carlo simulation (bullets). At the origin $\varphi$ is smooth though not visually fit.}
\label{fig3}
\end{figure} 
In order to calculate the coefficients and to understand their meaning it is convenient to express the eqs.~(\ref{bm-rs}) explicitly, thus
\begin{equation}
\lambda=4\,\int_{1-\varepsilon}^{+\infty}\varphi(x,x)\,\upd x + 2\,\int_{-\infty}^{1-\varepsilon}\varphi(1,x)\,\upd x\;.
\end{equation}
Here $\varepsilon\gtrsim 0$ is a small constant that avoids double evaluation of the integral at $x=1$.
One observes, that on one hand the coefficient $\lambda$ depends on integrals of $\varphi$ in points where the distribution $K$ is not smooth (see fig.~\ref{fig2}). On the other hand, the coefficient $\nu$ depends on the integral of the $\varphi$ where $K$ is not constant. Numerical simulations confirm these observations. In fact only the configurations $(j,j)$ for all $j\le 1$ and $(1,i)$ or $(i,1)$ for all $i<1$ contribute to $\lambda$, whereas other configurations of lateral growth are negligible. These conclusions were not reached by other approaches and could be confirmed by other unrestricted models with similar features, {\sl i.e.} discrete Langevin equation with not smooth drift functions. 

\paragraph*{Numerical results:}
The values of ​​SPDF obtained at points of configuration space of rules by Monte Carlo simulation in steady-state regime has allowed us to calculate the coefficients. The numerical values are $\lambda=0.3265$ and $v_0=2.1356$ which are in excellent agreement with those reported by Barabasi and Stanley \cite{Barabasi1995} using Monte Carlo simulation with initially tilted interface. Additionally, we calculated the value $\nu=1.0302$ which is not reported  by other authors.

\section{\label{sec:concl}Conclusions}
In this work we have proposed a new approach to calculate coarse grained coefficients and hence to determine the universality class of unrestricted and restricted discrete models. We present a theoretical framework which is based on well-established concepts of the distribution theory.

First we introduce the concept of a interface configuration space (ICS) where discrete processes occur. The analytic extension of ICS into real space allows us to define a test space. The stationary probability density function is embedded in the test function space which is generated by test functions with compact support or rapidly decreasing behaviour. The drift term of the discrete Langevin equation is understood as generalized function or distribution. We apply small constant translations in the test space and show how to obtain the coarse grained coefficients from the transformed average interface growth velocity.  The coefficients are determined by linear combinations of $n$-th order partial derivatives of a transformed drift distribution applied to the test function.

Then we study deposition models without relaxation, with and without restrictions, the RSOS and the BD models, that define spaces of test functions that are rapidly decreasing or with compact support, respectively. 
For these (1+1)-dimensional models, we determine analytically the coarse grained coefficients of KPZ class. 
The numerical values that we have determined for the nonlinear coefficient $\lambda$ are in very good agreement with those calculated by Monte Carlo simulation using the interface tilting method. 
In addition, our approach allows us to calculate the linear coefficient $\nu$ for these two models, which were not known until now. Apart from the pure deposition models we also study a competitive RSOS model that describes a restricted deposition-evaporation process. By this example we show how the symmetries of the distribution $K$ explain the crossover behaviour from KPZ universality class to EW ones. We observe that in all study cases $\lambda$ is expressed through integrals of the test function or its derivatives on the set of points where the $K$ distribution is not smooth or has edges or corners. 

Currently we have applied the formalism to volume conserving models \cite{Krug-97,Jung-99} which shows its full potential to determine the coefficients of conserved KPZ equations. These results will be reported shortly. 
For more complex lattice models, {\sl e.g.} see Ref.~\cite{Wolf-90}, the coefficients of the continuous equation will be dependent on the scale of observation. Recent works have calculated such coefficients by application of renormalization group \cite{Haselwandter-08, Haselwandter-10}.

\acknowledgments
We thank Sandra Molina (Mathematics Department at {FCEyN--UNMdP}) for her useful comments on Distributions Theory.  D.H. thanks CONICET for its support.

\bibliography{arxiv-Buceta-Hansmann-2012-final}

\end{document}